\documentstyle[preprint,aps]{revtex}
\begin{document}
\title{Quantum phase space distributions in thermofield dynamics} 
\author{S. Chaturvedi\thanks{e-mail:scsp@uohyd.ernet.in}, V. Srinivasan
\thanks {e-mail:vssp@uohyd.ernet.in}\\
School of Physics, University of Hyderabad, Hyderabad 500 046 India
\\
and
\\
G. S. Agarwal\thanks{email:gsa@prl.ernet.in}\\ Physical Research 
Laboratory, Navrangpura, Ahmedabad-380 009, India} 
\maketitle
\begin{abstract}
It is shown that the the quantum phase space distributions corresponding 
to a density operator $\rho$ can be expressed, in thermofield dynamics, 
as overlaps between the state $\mid \rho >$ and "thermal" coherent 
states. The usefulness of this approach is brought out in the context of  
a master equation describing a nonlinear oscillator for which exact 
expressions for the quantum phase distributions for an arbitrary initial 
condition are derived.  
\end{abstract}
\vskip1.5cm
\noindent PACS No: 03.65. Ca, 42.50.-{\bf p}  
\newpage
\section{Introduction}

In quantum mechanics, the state of a system is represented by a
density operator on a Hilbert space. The state can be pure or
mixed. To each density operator one can associate several
quantum phase space distributions which provide a quantum
analogue of the classical phase space distribution.  Prominent
among these are the Wigner function$^{1}$, the
$Q$-function$^{2}$ and the $P$-function$^{3}$ distribution
each with its own special features.  These three quantum phase
space distributions belong to a one parameter family of quantum
phase space distributions introduced and investigated by Cahill and 
Glauber$^{4}$ and by Agarwal and Wolf$^{5,6}$. Over the years,
quantum phase space distributions have not only proved to be
useful computational tools by enabling one to transcribe
operator equations into c-number language, but have also led to
new concepts such as non clasical states of radiation. Recent
developments in quantum state reconstruction$^{7}$, have made it
possible to measure some of the quantum phase space distributions
directly$^{8}$. They are no longer auxiliary concepts useful only for
computational purposes but have acquired a meaning in their own
right. Several schemes for direct measurements of quantum phase
space distributions$^{9}$ or the positivized versions thereof$^{10}$ 
as well as those in the context of atoms$^{11}$ have
been proposed and are likely to be experimentally realized in
the years to come. 

In this work, we examine the structure of the quantum phase
space distributions from the point of view of thermofield
dynamics$^{12,13}$.  In the conventional formulation of quantum mechanics,
pure states and mixed states are treated on an unequal footing.
The formalism of thermofield dynamics overcomes this drawback by
doubling the Hilbert space.  In this formalism, density
operators describing pure or mixed states are represented by a
state vector in the doubled Hilbert space.  The dynamical
equations in both cases also acquire the structure of a
Schr\"odinger equation even when dissipation is taken into
account.  The usefulness of this formalism for practical
purposes can be seen from the exact algebraic solution$^{14}$ of a
class of master equations$^{15}$ describing coupled dissipative
nonlinear oscillators.  It is shown here that, in the framework
of thermofield dynamics, quantum phase space distributions can
be expressed as overlap between the state corresponding to the
density operator and "thermal" coherent states.  To put it
differently, quantum phase space distributions appear as
coefficients of expansion whem the state vector corresponding to
a given density operator is expanded in terms of "thermal"
coherent states. That this aesthetically satisfying picture is
also useful is demonstrated by giving an exact and straightforward
 algebraic treatment for the evolution of the quantum phase space
distributions for the case of the master equation describing a
dissipative nonlinear oscillator.
   
A brief summary of this work is as follows. In sec. II we
briefly review the definition and properties of quantum phase
space distributions.  In sec.III we outline the formalism of
thermofield dynamics and show how the statements in sec I
translate into the thermofield dynamics notation. Some
illustrative examples of computation of quantum phase space
distributions are presented in sec. IV. Finally, in sec. V we
apply the machinery developed in the preceding sections to
obtain a complete picture of the time evolution of the phase
space distributions corresponding to the density operators which
evolve according to some standard master equations.
   
\section {Quantum phase space distributions}

In a series of papers Agarwal and Wolf have developed a general 
formulation of quantum phase space distributions. Crucial to their 
formulation is the notion of a $\Delta$-operator. Here we shall confine 
ourselves to  class of $\Delta$-operators relevant for the present work.
Consider the family of operators 
\begin{equation}
\Delta^{(a)}(\alpha,\alpha^* ) = 
\frac{1}{\pi} 
\int d^2 \beta D(\beta)e^{({\it a}-\frac{1}{2}){\mid\beta\mid}^2
-(\beta\alpha^{*}-\beta^{*} \alpha)}~~~;~~~{\it a}~\leq~1~~,
\end{equation}
where $D(\alpha) = \exp(\alpha a^\dagger -\alpha^{*} a)$. These operators 
are hermitian 
\begin{equation}
\Delta^{(a)\dagger}(\alpha,\alpha^* ) = \Delta^{(a)}(\alpha,\alpha^* )~~, 
\end{equation}
and have the following properties
\begin{equation}
Tr [\Delta^{(a)}(\alpha,\alpha^* )] = \pi \delta^2 (\alpha) ~~,
\end{equation}
\begin{equation}
Tr [\Delta^{(a)}(\alpha,\alpha^* )\Delta^{(1-a)}(\beta,\beta^* )] = 
\pi \delta^2 (\alpha-\beta)~~,
\end{equation}
which follow from the fact that $Tr[D^\dagger (\alpha) D(\beta) = \pi 
\delta^2 (\alpha -\beta)$. With the help of these operators one can 
associate with a density operator $\rho$
a class of quantum phase space distributions as follows
\begin{equation}
\Phi^{({\it a})}_\rho (\alpha, \alpha^{*}) = \frac{1}{\pi } 
Tr[\rho \Delta^{(a)}(\alpha,\alpha^* )]~~~.
\end{equation}
Conversely, $\rho$ can be expressed in terms of these in the following 
manner
\begin{equation}
\rho = \int d^2 \alpha \Phi^{({\it a})}_\rho (\alpha, \alpha^{*})
 \Delta^{(1-a)}(\alpha,\alpha^* )~~~.
\end{equation}
For ${\it a} = 1,~1/2,~0$, $\Phi^{({\it a})}_\rho (\alpha, \alpha^{*})$  
respectively correspond to the $P$-function $P(\alpha,\alpha^{*})$, the Wigner 
function $W(\alpha,\alpha^{*})$ and the $Q$-function $Q(\alpha,\alpha^{*})$.
The family of quantum phase space distributions defined in this way 
are related to each other as follows 
\begin{equation}
\Phi^{({\it a-b})}_\rho (\alpha, \alpha^{*})
=\frac{1}{\pi {\it b}}\int d^2 \beta 
\Phi^{({\it a})}_\rho (\beta, \beta^{*}) 
\exp \left[ -\frac{{\mid \alpha -\beta \mid}^2}{{\it b}}\right]~;~
 {\it b}~\leq~{\it a}~~~,
\end{equation}
\begin{equation}
\Phi^{({\it a})}_\rho (\alpha, \alpha^{*}) = 
\exp\left[ -({\it a-b}) \frac{\partial^2}{\partial \alpha 
\partial \alpha^{*}}\right]
\Phi^{({\it b})}_\rho (\alpha, \alpha^{*})~~~.
\end{equation}
It may be noted that the class of $\Delta$-operators may also be expressed 
in the following manner
\begin{equation}
\Delta^{(a)}(\alpha,\alpha^* ) = D(\alpha)\rho_{0}^{(a)} D^\dagger (\alpha)~~,
\end{equation}
where
\begin{equation}
\rho_{0}^{(a)} = \frac{1}{\pi} 
\int d^2 \beta D(\beta)e^{({\it a}-\frac{1}{2}){\mid\beta\mid}^2}~~~.
\end{equation}
The operator $\rho_0$ can be expressed in a more familiar form as 
\begin{equation}
\rho_{0}^{(a)} = (1-e^{-\theta}) e^{-\theta a^\dagger a} ~~~; 
~~~~e^{-\theta} = -\frac{a}{1-a}~~~, 
\end{equation}
and is easily seen to have the structure of a "thermal" vacuum. It 
correspnds to a genuine thermal vacuum only for $a \leq 0$. The 
family of quantum phase space distributions can therefore be compactly 
defined as follows  
\begin{equation}
\Phi^{({\it a})}_\rho (\alpha, \alpha^{*})= \frac{1}{\pi} 
Tr[\rho D(\alpha)\rho_{0}^{(a)} D^\dagger (\alpha)]~~~.
\end{equation}
This observation will be crucial to later developments. It may be remarked 
here that, expressed in this way, the quantum phase space distributions 
corresponding to a density operator $\rho$ are identifiable as the generating 
function of the number distribution of the density operator 
$ D^\dagger (\alpha) \rho D(\alpha)$.
\begin{equation}
\Phi^{({\it a})}_\rho (\alpha, \alpha^{*})
= \frac{1-\lambda}{\pi}\sum_{n} {\lambda}^n <n\mid   
D^\dagger (\alpha) \rho D(\alpha)\mid n>~~~~;~~~~~ \lambda = -a/(1-a)~~.
\end{equation}
That the family of quantum phase space distributions 
$\Phi^{({\it a})}_\rho (\alpha, \alpha^{*})$ can be expressed as in $(13)$, 
was, to our knowledge, first noticed by Moya-Cessa and Knight$^{16}$. 
 
\section{Thermofield dynamics}

In TFD one associates, with a density operator $\rho$ acting on  a 
Hilbert space ${\cal H},a$ state vector  $\mid \rho^\alpha>,  1/2  \le 
\alpha \le 1$ in the extended Hilbert space ${\cal H}\otimes {\cal 
H}^*$ so that averages of operators with respect to $\rho$ acquire 
the appearance of a scalar product.
\begin{equation}
<A> = Tr A\rho = <\rho^{1-\alpha}\mid A\mid \rho^\alpha>\,\,\,.
\end{equation}
The state $\mid \rho^\alpha>$ is given by
\begin{equation}
\mid \rho^\alpha> = \hat{\rho}^\alpha\mid I>\,\,\,\,,
\end{equation}
where
\begin{equation}
\hat{\rho}^\alpha = \rho^\alpha \otimes I\,\,\,\,,
\end{equation}
and
\begin{equation}
\mid I> = \sum \mid N>\otimes\mid N> \equiv \sum \mid N,N>\,\,\,,
\end{equation}
where $\mid N>$ constitute any complete orthonormal set in ${\cal H}$. 
The state $\mid I>$ is simply the counterpart of the resolution of the 
identity
\begin{equation}
\mid I> = \sum \mid N><N\mid \,\,\,,
\end{equation}
in terms of a complete orthonormal set  $\mid N>$  in  ${\cal  H}$.  In 
particular, if
\begin{eqnarray}
\rho\mid N> & = & p_N\mid N> \,\,\,,\\
{}\mid \rho^\alpha> & = & \sum_N p_N^\alpha \mid N,N>\,\,\,\,.
\end{eqnarray}

(It may  be  noted  that  for  any  density  operator  the  states 
$\mid \rho^\alpha>, 1/2 \le \alpha \le1$ have a  finite  norm  in  the 
extended Hilbert space. This is not necessarily so with the states 
$\mid \rho^{1-\alpha}>, 1/2 \le \alpha \le1$, which include the  state 
$\mid I>$. These states are to be regarded  as  formal  but  extremely 
useful constructs)

In this work we shall set $\alpha=1$. This representation of the density 
operator as a state vector in the extended Hilbert space is the only useful 
one for discussing dissipative dynamics which will be of interest to us later. 
In this representation, for any hermitian operator $A$,one has 
\begin{equation}
<A> = Tr(A\rho) = <A\mid\rho>~~.
\end{equation} 
Further, for  $\mid N>$, we choose 
the number state $\mid n>$ and introduce creation and  annihilation 
operators $a^\dagger, \tilde{a}^\dagger,  a$,  and  $\tilde{a}$  as 
follows
\begin{eqnarray}
a\mid n,m> = \sqrt{n} \mid n-1,m> ~~~~~ &,&~~~~~ \tilde{a}\mid n,m> =  
\sqrt{m} \mid n,m-1>,\\
a^\dagger\mid n,m> = \sqrt{n+1} \mid n+1,m>~~&,&~~ \tilde{a}^\dagger\mid n,m> 
 = \sqrt{m+1} \mid n,m+1>.
\end{eqnarray}
The operators $a$ and $a^\dagger$  commute  with  $\tilde{a}$  and 
$\tilde{a}^\dagger$.  It  is  easily  seen  that   the   operators 
$\tilde{a}$  and  $\tilde{a}^\dagger$  respectively  simulate   the 
action of $a^\dagger$ and $a$ on $\mid n><m\mid $ from the right. From the 
expression for $\mid I>$ in terms of the number states
\begin{equation}
\mid I> = \sum_n \mid n,n>\,\,\,\,,
\end{equation}
it follows that
\begin{equation}
a\mid I>   =   \tilde{a}^\dagger    \mid I>\,\,\,\,\,\,; a^\dagger\mid I>    = 
\tilde{a}\mid I>\,\,\,\,,
\end{equation}
and hence for any operator
\begin{equation}
A(a^\dagger,  a)   =   \sum_{p,q}   \alpha_{p,q}   a^{\dagger   p} 
a^q\,\,\,\,,
\end{equation}
one has
\begin{equation}
A\mid I> = \tilde{A}^\dagger \mid I>\,\,\,\,,
\end{equation}
where $\tilde{A}$ is obtained from $A$ by making the  replacements 
(tilde  conjugation  rules)   $a\to   \tilde{a},   a^\dagger   \to 
\tilde{a}^\dagger,  \alpha\to  \alpha^*$.

With the help of the formal constructs introduced above, it becomes possible 
to represent any density operator $\rho$ as a state vector in the extended 
Hilbert space. Thus, for instance, the density operator for a coherent state 
$\rho= \mid\alpha><\alpha\mid$ can be represented as $\mid\rho> = 
D(\alpha)\tilde{D}(\alpha^* )\mid 0,0>$. Similarly for the density operator 
corresponding to a thermal state one has 
\begin{equation}
\rho =(1-e^{-\beta})e^{-\beta a^\dagger a}~~~ \rightarrow~~~
\mid\rho> = (1-f)e^{fK_+}\mid 0,0>~~~,
\end{equation} 
where $f= e^{-\beta}$ and $K_+ = a^\dagger{\tilde{a}}^\dagger$.  The operator 
$K_+ $ together with $K_- = a\tilde{a}$ and $K_3 = (a^\dagger a + {\tilde{a}}^
\dagger \tilde{a} +1)/2$ satisfies the algebra of $su(1,1)$. 
\begin{equation}
[K_-, K_+] = 2K_3\,\,\,;\,\,\,\, [K_3, K_\pm] = \pm K_\pm\,\,\,\,,
\end{equation}
with $K_o = (a^\dagger a - {\tilde{a}}^
\dagger \tilde{a})$ as the Casimir operator. Use of the disentangling theorem 
for $su(1,1)^{14}$
\begin{equation}
\exp(\gamma_+K_++\gamma_3K_3+\gamma_-K_-) = \exp(\Gamma_+K_+) 
\exp((2\log \sqrt{\Gamma_3)}K_3) \exp(\Gamma_-K_-)\,\,\,,
\end{equation}
where
\begin{equation}
\Gamma_\pm    =     {2\gamma_\pm\sinh\phi\over2\phi\cosh\phi     - 
\gamma_3\sinh\phi}\,\,\,,\,\,\,\sqrt{\Gamma_3}  =  \left({2\phi\over2\phi 
\cosh\phi - \gamma_3 \sinh\phi}\right)\,\,\,,
\end{equation}
with
\begin{equation}
\phi^2 = (\gamma^2_3/4) - \gamma_+\gamma_-\,\,\,\,,
\end{equation}
enable us to write the state $\mid \rho>$ in (28) as  
\begin{equation}
\mid\rho> = e^{\bar{n}(K_+ + K_- -2K_3}\mid 0,0>~~~.
\end{equation}

Turning to dynamics, consider, for example, the master equation for a 
nonlinear oscillator 
\begin{equation}
{\partial\over\partial t}     \rho     =     -i[H, \rho]      + 
{1\over2}\gamma   (\bar{n}+1)(2a\rho   a^\dagger    -    a^\dagger 
a\rho-\rho a^\dagger a) + {1\over2}\gamma\bar{n}(2a^\dagger\rho  a 
- aa^\dagger\rho - \rho a a^\dagger)\,\,,
\end{equation}
where $H=\omega  a^\dagger  a+\chi(a^\dagger 
a)^2$. This master equation has been studied by a number of authors$^{15}$ 
in the context of nonlinear propagation in a Kerr medium. Applying $\mid I>$ 
on  (34)  from  the 
right and using (25) this master equation for $\rho$ goes over to a 
Schr\"odinger like equation for the state $\mid \rho>$ 
\begin{equation}
{\partial\over\partial t} \mid \rho>  = -i\hat{H}\mid \rho>\,\,\,,
\end{equation}
where
\begin{eqnarray}
-i\hat{H} = &-&i\omega(a^\dagger a-\tilde{a}^\dagger\tilde{a}) 
-i\chi [(a^\dagger a)^2-(\tilde{a}^\dagger\tilde{a})^2]  
+  {1\over2}   \gamma(\bar{n}+1)(2a\tilde{a}-a^\dagger 
a-\tilde{a}^\dagger   \tilde{a}) \nonumber \\  &+&    
{1\over2}\gamma \bar{n}(2a^\dagger 
\tilde{a}^\dagger     -      aa^\dagger-\tilde{a} 
\tilde{a}^\dagger)\,\,\,\,.
\end{eqnarray}
In terms of the operators $K_+, K_-, K_3$ and $K_0$, the operator $-i\hat{H} $ 
can be written as 
\begin{equation}
-i\hat{H}     =     -i(\omega-\chi)K_o+\gamma(\bar{n}+1)K_-+\gamma 
\bar{n}K_+   -   (\gamma(2\bar{n}+1)+2i\chi   K_o)   K_3+{1\over2} 
\gamma\,\,\,\,,
\end{equation}
and hence the solution of (35) as
\begin{equation}
\mid \rho(t)> =\exp(\gamma_0 K_0 + {1\over2} \gamma t) \exp(\gamma_+K_+ 
 +\gamma_3K_3 + \gamma_-K_-)\mid \rho(0)>\,\,\,\,,
\end{equation}
where 
\begin{equation}
\gamma_+ = \gamma\bar{n}t \,\,; \gamma_-=\gamma(\bar{n}+1)t \,\,;  \gamma_3 
= -(\gamma(2\bar{n}+1) +2i\chi K_o)t \,\,; \gamma_o = -i(\omega-\chi) t
\,\,\,\,.
\end{equation}
Using the disentangling theorem (30), (38) can be written as 
\begin{equation}
\mid \rho(t)> =\exp(\gamma_0 K_0 + {1\over2} \gamma t) 
\exp(\Gamma_+K_+) 
\exp((2\log \sqrt{\Gamma_3)}K_3) \exp(\Gamma_-K_-)\mid \rho(0)>\,\,\,\,,
\end{equation}
The fact that $K_+, K_-$ and $K_3$ have simple  actions  on 
$\mid n,m>$  enable  one  to  solve  (35)  and   hence   (34)   purely 
algebraically. Detailed expressions for  $\rho_{m,n}(t)$  and  the 
$Q$-function for an arbitrary initial condition may  be  found  in 
ref.14.

In the context of the interaction of a single field mode with a nonthermal 
and phase insensitive environment, Agarwal$^{18}$ has considered the following 
master equation
\begin{equation}
\frac{\partial}{\partial t} \rho = \kappa[a \rho a^{\dagger} + a^{\dagger} 
\rho a 
-(a^{\dagger} a + \frac{1}{2})\rho - \rho (a^{\dagger}a + \frac{1}{2})] ~~,
\end{equation}
In thermofield dynamics notation this translates into 
\begin{equation}
\frac{\partial}{\partial t} \mid \rho> = \kappa[K_+ +K_- - 2K_3]\mid \rho>~,
\end{equation}
so that
\begin{equation}
\mid \rho(t)> = \exp(\kappa t(K_+ + K_- -2K_3))\mid \rho(0)>.
\end{equation}

To conclude this section, we have seen how various density density operators 
are represented in thermofield dynamics. We have also seen how certain master 
equations could be solved purely algebraically using the thermofield dynamics 
formalism. In the following section we shall show how the quantum phase space 
distributions are to be calculated in thermofield dynamics and how the time 
evolution of the quantum phase space distributions corresponding to the 
the master equations considered here can be  exactly determined.
\section {Quantum phase space distributions in thermofield dynamics}

From (12) and (21) it follows that
\begin{equation}
\Phi^{({\it a})}_\rho (\alpha, \alpha^{*}) = 
\frac{1}{\pi}<\alpha,\alpha^{*};{\it a}\mid \rho>~, 
\end{equation}
where 
\begin{eqnarray}
\mid \alpha,\alpha^{*};{\it a}> &=&  
\exp\left[ -{\it a}[K_{+} + K_{-} -2 K_z]\right] \mid \alpha, \alpha^{*}>,
\\
&=& \frac{1}{1-{\it a}} D(\alpha)\tilde{D}(\alpha^{*})
\exp\left[ -\frac{{\it a}}{1-{\it a}} K_+\right] \mid 0, 0>~~.
\end{eqnarray}
The quantum phase space distributions can thus be written as overlaps 
between the state $\mid \rho>$ and "thermal" coherent states. We now consider 
some examples to illustrate the usefulness of this formula for computation 
of quantum phase space distributions for some standard density operators.

\noindent 
[1] Coherent states
 
Consider the density operator $\rho = \mid \alpha_0><\alpha_0\mid$ for a 
coherent state. The corresponding state 
$\mid \rho>$ is  $\mid \alpha_0, \alpha_{0}^{*}>$. Substituting this in (44) 
we obtain
\begin{eqnarray}
\Phi^{({\it a})}_\rho (\alpha, \alpha^{*}) &=& 
\frac{1}{(1-{\it a})\pi} <0,0\mid \exp\left[ -\frac{{\it a}}{1-{\it a}} K_
-\right]
D(-\alpha)\tilde{D}(-\alpha^{*})
\mid \alpha_0, \alpha_{0}^{*}>~~, 
\nonumber \\
&=& \frac{1}{(1-{\it a})\pi}
<0,0\mid \exp\left[ -\frac{{\it a}}{1-{\it a}} K_-\right]
\mid \alpha_0-\alpha, \alpha_{0}^{*}-\alpha^{*}> \nonumber\\
&=&\frac{1}{(1-{\it a})\pi}\exp\left[ -\frac{{\it a}}{1-{\it a}} 
\mid \alpha-\alpha_0\mid^2 \right]
<0,0\mid \alpha_0-\alpha, \alpha_{0}^{*}-\alpha^{*}>~~, \nonumber\\
&=&\frac{1}{(1-{\it a})\pi}\exp\left[ -\frac{1}{1-{\it a}} 
\mid \alpha-\alpha_0\mid^2 \right]~~.
\end{eqnarray}
In the limit $a\rightarrow 1$ one obtains the familiar expression for the 
P-function for the coherent state
\begin{equation} 
\Phi^{({\it 1})}_\rho (\alpha, \alpha^{*}) = 
\frac{1}{\pi}<\alpha,\alpha^{*};{\it 1}\mid \rho> = \delta^2 (\alpha-\alpha_0)
~. 
\end{equation}
Now, since 
\begin{equation}
<\alpha,\alpha^{*};{\it a}\mid \alpha_0,\alpha_{0}^{*}; b>
=<\alpha,\alpha^{*};{\it a+b}\mid \alpha_0,\alpha_{0}^{*}>~~,
\end{equation}
it follows that 
\begin{equation}
<\alpha,\alpha^{*};{\it a}\mid \alpha_0,\alpha_{0}^{*}; b>~= 
\begin{array}{c}~~~~ \frac{1}{(1-({\it a+b}))}
\exp\left[ -\frac{1}{1-({\it a+b})} 
\mid \alpha-\alpha_0\mid^2 \right] ~~~if~~~ {\it a+b}~< 1 \\ ~~~~~~~~~~\pi 
\delta^2 (\alpha-\alpha_0)~~~~~~~~~~~~~~~~~~~~~~~~~~ if~~~ {\it a+b}~
=~1\end{array}
\end{equation}
This relation will be used later.

\noindent
[2] Squeezed thermal coherent state$^{19}$

The density operator for a squeezed thermal coherent state is 
\begin{equation}
\rho = (1-e^{-\beta}) D(\alpha_0)S(z)\exp (-\beta a^\dagger a)
S^\dagger(z)D^\dagger(\alpha_0)~~.
\end{equation}
The corresponding $\mid\rho>$ is given by
\begin{equation}
\mid \rho> = (1-f) D(\alpha_0)\tilde{D}(\alpha_{0}^{*})S(z)
\tilde{S}(z^{*}) \exp \left[ fK_+ \right]\mid 0, 0>
~~;~~f~=~e^{-\beta}~~~.
\end{equation}
Substituting this in (44) we obtain
\begin{equation}
\Phi^{({\it a})}_\rho (\alpha, \alpha^{*})
=\frac{(1-f)(1-\lambda)}{\pi}
<0,0\mid e^{\lambda K_-}D^\dagger(\alpha)\tilde{D}^\dagger (\alpha^*)
D(\alpha_0)\tilde{D}(\alpha_{0}^{*})S(z)\tilde{S}(z^*)e^{fK_+}\mid 0,0>. 
\end{equation}
By inserting the resolution of the identity in terms of two mode coherent 
state after $\exp (\lambda K_-)$ and before $\exp(fK_+)$, writing out the 
action of these operators on the coherent states, and carrying out two 
of the four integrals that occur we obtain
\begin{eqnarray}
\Phi^{({\it a})}_\rho (\alpha, \alpha^{*})=
\frac{(1-f)(1-\lambda)}{\pi}\int & & \frac{d^2\gamma}{\pi}
\int \frac{d^2\delta}{\pi}e^{-\frac{1}{2}(1-\lambda^2)\mid\gamma\mid^2
 +\frac{1}{2}(1-\lambda)(\gamma\alpha^*-\gamma^*\alpha) 
-\frac{1}{2}(1-f^2)\mid\delta\mid^2} \nonumber\\
& &<\gamma+\alpha-\alpha_0,\lambda\gamma^* +\alpha^*-\alpha_{0}^{*}
\mid S(z)\tilde{S}(z*)\mid\delta,f\delta^* >.
\end{eqnarray}
Finally, on using the known expression for the matrix element of the 
squeezing operator $S(z)$ between coherent states and carrying out the 
gaussian integrals, we obtain  
\begin{equation}
\Phi^{({\it a})}_\rho (\alpha, \alpha^{*})=
\frac{(1-f)(1-\lambda)sechr}{\pi\sqrt{[(1-\lambda f)^2 -(\lambda-f)^2 
tanh^2 r]}}exp[-\frac{(1-f)(1-\lambda)}{[(1-\lambda f)^2 -
(\lambda-f)^2 tanh^2 r]}X]~,
\end{equation}
where $\lambda = -a/(1-a)$ $ z= re^{i\theta}$ and
\begin{eqnarray}
X &=& [(1-\lambda f)-(\lambda-f)tanh^2 r]
\mid\alpha-\alpha_0\mid^2\nonumber\\
& & -\frac{1}{2}(1+f)(1-\lambda)tanhr
[(\alpha-\alpha_0)^2 e^{-i\theta} 
+(\alpha^{*}-\alpha_{0}^{*})^2e^{i\theta}]~.
\end{eqnarray}
In the limit of no squeezing, one recovers the familiar 
expressions for the quantum phase space distributions for a thermal 
coherent state
\begin{equation}
\Phi^{({\it a})}_\rho (\alpha, \alpha^{*})  
= \frac{(1-f)}{(1-{\it a}(1-f))\pi}
\exp \left[ \frac{ (1-f)}{1-{\it a}(1-f)} {\mid \alpha-\alpha_0 \mid}^2\right].
\end{equation}

With this preparation, we are in a position to translate all the relations in 
sec. II into the language of thermofield dynamics 

An arbitrary state $\mid\rho >$ can be in terms 
of $\mid \alpha,\alpha^{*};{\it a}>$ as follows
\begin{equation}
\mid \rho> = \int d^2 \beta \Phi^{({\it a})}_\rho (\beta, \beta^{*})
\mid \beta,\beta^{*};{\it 1-a}>~.
\end{equation}
This relation is the analogue of (6). The quantum phase space distribution 
thus acquire a new meaning as  coefficients of expansion when $\mid\rho>$ is 
expressed in terms of the "thermal" coherent states 
$\mid \alpha,\alpha^{*};{\it 1-a}> $. 

Taking the overlap of $\mid\rho>$ as given in (58) with 
$\mid \alpha,\alpha^{*};{\it a-b}>$ and using (49) and (50) one obtains the 
relation $(2)$ between quantum phase space distributions. 

The relation (8) between the quantum phase space distributions comes 
about in the following way. The coherent state $\mid \alpha, \alpha^* >$ 
has the following structure 
\begin{equation}
\mid \alpha, \alpha^* > =\exp (\alpha(a^\dagger-\tilde{a}) -\alpha^* (a-
{\tilde{a}}^\dagger))\mid 0,0>~.
\end{equation}
The operators $(a^\dagger-\tilde{a})$ and $(a-{\tilde{a}}^\dagger))$ commute 
with each other. Now, from (59) it follows that  
\begin{equation}
(a^\dagger-\tilde{a}) \mid \alpha, \alpha^* > = 
\frac{\partial}{\partial \alpha}\mid \alpha, \alpha^* > ~~;~~  
(a-{\tilde{a}}^\dagger))\mid \alpha, \alpha^* >=  
\frac{\partial}{\partial \alpha^* }\mid \alpha, \alpha^* >~~,
\end{equation}
and hence 
\begin{equation} 
[K_+ + K_- -2K_3] \mid \alpha, \alpha^* > = -
(a^\dagger-\tilde{a})(a-{\tilde{a}}^\dagger))\mid \alpha, \alpha^* > =- 
\frac{\partial^2}{\partial\alpha \partial \alpha^*} \mid \alpha, \alpha^* >.
\end{equation} 
The relation (8) therefore is a simple consequence of (44) and of the 
realization of $[K_+ + K_- -2K_3]$ as a differential operator while 
acting on $\mid \alpha, \alpha^* >$.  

Finally, the representation (58) also leads to the relation 
\begin{equation}
Tr(A\rho) =\pi \int d^2 \beta 
\Phi^{({\it a})}_\rho (\beta, \beta^{*}) 
\Phi^{({\it 1-a})}_A (\beta, \beta^{*}).  
\end{equation}

\section { Time evolution of quantum phase space distributions:}
In sec. III we discussed some standard master equations and showed how they 
could be transcribed as Schr\"odinger like equations. In the cases 
considered we showed that the solution of the Schr\"odinger like 
equations thus obtained can be written as (38) with appropriate 
identification of the coefficients that occur there. In this section we 
wish to examine how transcription enables us to obtain a complete picture 
of the corresponding phase space distributions subject to given initial 
conditions. Two initial conditions are considered 

\noindent
[1] Evolution from an initial coherent state 
$\mid\rho(0)> = \mid \alpha_0,\alpha_{0}^{*}>$

For this initial condition, it follows from (38), (44) and (45) that 
\begin{eqnarray}
& &\Phi^{({\it a})}_{\rho(t)} (\alpha, \alpha^{*}) \nonumber\\ 
&=&\frac{1}{\pi}<\alpha,\alpha^{*};a\mid
e^{( \gamma_0 K_0+\frac{1}{2}\gamma t)}
e^{(\Gamma_+K_+)}e^{((2\log\sqrt{\Gamma_3})K_3)}
e^{(\Gamma_-K_-)}\mid\alpha_0,\alpha_{0}^{*}> 
\nonumber\\
&=& \frac{1}{\pi}e^{\frac{1}{2} \gamma t}<\alpha,\alpha^{*}\mid
e^{(-a(K_++ K_-- 2K_3)}
e^{(\Gamma_+K_+)}e^{((2\log\sqrt{\Gamma_3})K_3)}
e^{(\Gamma_-K_-)}
\alpha_0e^{-i\omega t},\alpha_{0}^{*}e^{i\omega t}>\nonumber\\
&=&\frac{1}{\pi}e^{\frac{1}{2} \gamma t}<\alpha,\alpha^{*}\mid 
e^{(\lambda K_+)}e^{((2\log (1-\lambda)K_3)}e^{(\lambda K_-)}
e^{(\Gamma_+K_+)} 
e^{((2\log \sqrt{\Gamma_3})K_3)} e^{(\Gamma_-K_-)}\mid 
\alpha_0e^{-i\omega t},\alpha_{0}^{*}
e^{i\omega t}>,\nonumber\\
\end{eqnarray}
where $\lambda = -a/(1-a)$. Here in the last step we have used the 
disentangling theorem (30). Repeated use of the commutation relations 
and the disentangling theorem enable us to bring the rhs of (63) into the 
following standard form
\begin{equation}
\Phi^{({\it a})}_{\rho(t)} (\alpha, \alpha^{*})=
\frac{1}{\pi}<\alpha,\alpha^{*}\mid 
e^{(\Gamma_{+}^{\prime} K_+)} 
e^{((2\log \sqrt{\Gamma_{3}^{\prime}})K_3)} 
e^{(\Gamma_{-}^{\prime} K_-)}\mid\alpha_0e^{-i\omega t},\alpha_{0}^{*}
e^{i\omega t}>~,
\end{equation}
where
\begin{eqnarray}
\Gamma_{+}^{\prime} &=& 1 -\frac{[(1-\lambda)(1-\Gamma_+)]}{[1-\lambda 
\Gamma_-]},
\nonumber\\ \Gamma_{-}^{\prime} &=& 
1 - \frac{[(1-\Gamma_-)(1-\lambda\Gamma_+)-\lambda 
\Gamma_3]}{[1-\lambda\Gamma_+]},\\
\sqrt{\Gamma_{3}^{\prime}} &=&
\frac{[(1-\lambda)\sqrt{\Gamma_3}]}{[1-\lambda\Gamma_+]}~.\nonumber
\end{eqnarray} 
In this form the rhs of (64) can easily be evaluated. 
In the case of the nonlinear oscillator 
$\Gamma_{\pm}$ and $\Gamma_3$ are 
functions of $K_0$ and (64) can be written as 
\begin{equation}
\Phi^{({\it a})}_{\rho(t)} (\alpha, \alpha^{*})
=\frac{1}{\pi}\sum_{m,n} \frac{1}{m!n!} 
(\alpha^* \alpha_0 e^{-i\omega t})^m
 (\alpha \alpha_{0}^{*} e^{i\omega t})^n
(\sqrt{\Gamma_3})^{m+n+1}e^{(\Gamma_{+}^{\prime}-1){\mid \alpha\mid}^2}
e^{(\Gamma_{-}^{\prime}-1){\mid \alpha_0\mid}^2}~.
\end{equation}
where it is understood that the $K_o$ in the expressions for 
$\Gamma_{\pm}$ and $\Gamma_3$ is replaced by $(m-n)$.

 When $\chi =0)$, i.e. in the case of a linear oscillator, 
$\Gamma_{\pm}$ and $\Gamma_3$ are constants and (66) simplifies to  
\begin{equation}
\Phi^{({\it a})}_{\rho(t)} (\alpha, \alpha^{*})= 
\frac{\sqrt{\Gamma_{3}^{\prime}}}{\pi}e^{\frac{1}{2} \gamma t}
e^{(\Gamma_{+}^{\prime}-1){\mid \alpha\mid}^2}
e^{(\Gamma_{-}^{\prime}-1){\mid \alpha_0\mid}^2}
e^{\sqrt{\Gamma_{3}^{\prime}}(\alpha^* \alpha_0 e^{-i\omega t}
+\alpha \alpha_{0}^{*} e^{i\omega t})}~.
\end{equation}

\noindent
[2] Evolution from a given initial quantum phase space distribution

An arbitrary initial state can be expanded as 
\begin{equation}
\mid \rho(0)> = \int d^2 \alpha_0 \Phi^{({\it a})}_{\rho(0)} (\alpha_0, 
\alpha_{0}^{*})
\mid \alpha_{0},\alpha_{0}^{*};{\it 1-a}>~.
\end{equation}
For the time evolution of the quantum phase space distributions one 
therefore obtains 
\begin{equation}
\Phi^{({\it a})}_{\rho(t)} (\alpha, \alpha^{*})=
\int d^2\alpha_0 K^{(a)}(\alpha,\alpha^{*},t~;~\alpha_0,\alpha_{0}^{*},0)
\Phi^{({\it a})}_{\rho(0)} (\alpha_0, \alpha_{0}^{*})~,
\end{equation}
where 
\begin{equation}
K^{(a)}(\alpha,\alpha^{*},t~;~\alpha_0, \alpha_{0}^{*},0)
=\frac{1}{\pi}<\alpha,\alpha^{*};a\mid
e^{( \gamma_0 K_0+\frac{1}{2}\gamma t)}
e^{(\gamma_+K_++\gamma_3K_3+\gamma_-K_-)}\mid\alpha_0,\alpha_{0}^{*}; (1-a)>. 
\end{equation}
Using the same manipulations as above one finds that
\begin{equation}
K^{(a)}(\alpha,\alpha^{*},t~;~\alpha_0, \alpha_{0}^{*},0)= 
\frac{e^{\frac{\gamma t}{2}}}{\pi}<\alpha,\alpha^{*}\mid 
e^{(\Gamma_{+}^{\prime\prime} K_+)} 
e^{((2\log \sqrt{\Gamma_{3}^{\prime\prime}})K_3)} 
e^{(\Gamma_{-}^{\prime\prime} K_-)}\mid\alpha_0e^{-i\omega t},\alpha_{0}^{*}
e^{i\omega t}>,
\end{equation}
where
\begin{eqnarray}
\Gamma_{+}^{\prime\prime}&=& 1 - (1-\lambda)\frac{[(1-\Gamma_+)(\lambda-
\Gamma_-)-\Gamma_3]}{[(\lambda-\Gamma_-)(1-\lambda\Gamma_+)-\lambda\Gamma_3]} 
,\nonumber\\
\Gamma_{-}^{\prime\prime}&=& 1 + (1-\lambda)\frac{[(1-\Gamma_-)(1-\lambda
\Gamma_+)-\lambda\Gamma_3]}{[(\lambda-\Gamma_-)(1-\lambda\Gamma_+)-\lambda 
\Gamma_3]},\\
\sqrt{\Gamma_{3}^{\prime\prime}}&=& -\frac{(1-\lambda)^2\sqrt{\Gamma_3}}
{[(\lambda-\Gamma_-)(1-\lambda\Gamma_+)-\lambda\Gamma_3]}.\nonumber
\end{eqnarray}

The rhs of (71) can be cast into the form (67) for the case the linear 
oscillator and in the form (66) in the case of the nonlinear oscillator.

Finally, for the master equation (41), owing to its structure, one has the 
following interesting result
\begin{equation}
\Phi^{({\it a})}_{\rho(t)} (\alpha, \alpha^{*}) = 
\Phi^{({\it a-\kappa t})}_{\rho(0)} (\alpha, \alpha^{*})~, 
\end{equation}
which says that $\Phi^{({\it a})}_{\rho(t)} (\alpha, \alpha^{*})$ sweeps 
through the entire family of the quantum phase space distributions 
$\Phi^{({\it a})}_{\rho(t)} (\alpha, \alpha^{*})~~ ;~~b~\leq ~a$
associated with the initial density operator
\vskip3.0cm 
\section{Conclusion}

The observation that the family of quantum phase space distributions 
considered by Agarwal and Wolf have the structure as in (12) and the 
recognition that this form, in thermofield dynamics, translates into 
an overlap as in (44) constitute the key results of this work. These 
results, in turn, enable us not only to understand various relations 
between quantum phase space distribution in an aesthetically statifying 
manner, but also prove to be extremely useful for computational purposes. 
The latter aspect of the formalism developed here is demonstrated by 
using it to obtain  a full time dependent solution for a family of 
quantum phase space distributions for the master equation describing a 
nonlinear dissipative oscillator. We hope that the results presented here 
will be 
useful in the context of wave packet dynamics in nonlinear systems 
in the presence of dissipation, a subject, which has attracted considerable 
attention in recent years.

\end{document}